\newcommand{\diracslash}[1]{#1\llap{/\kern2pt}}

\newcommand{\be}{\begin{equation}}
\newcommand{\ee}{\end{equation}}
\newcommand{\bea}{\begin{eqnarray}}
\newcommand{\eea}{\end{eqnarray}}
\newcommand{\ba}[1]{\begin{array}{#1}}
\newcommand{\ea}{\end{array}}

\documentclass[prd,aps,floats,nofootinbib,tightenlines,showpacs]{revtex4}
\usepackage{epsfig,graphicx,pstricks}
\usepackage{psfrag}
\usepackage{color}
\usepackage{amsmath}
\usepackage{amsfonts}
\usepackage{amssymb}
\usepackage{textcomp}
\usepackage{multirow}
\usepackage{subfigure}
\usepackage{slashed}

\begin{document}

\title {On the thermodynamically consistent quasiparticle model of quark gluon plasma}
\author{Guru Prakash Kadam }
\email{guruprasad@prl.res.in}
\affiliation{Theory Division, Physical Research Laboratory,
Navrangpura, Ahmedabad 380 009, India}

\date{\today} 

\def\be{\begin{equation}}
\def\ee{\end{equation}}
\def\bearr{\begin{eqnarray}}
\def\eearr{\end{eqnarray}}
\def\zbf#1{{\bf {#1}}}
\def\bfm#1{\mbox{\boldmath $#1$}}
\def\hf{\frac{1}{2}}
\def\sl{\hspace{-0.15cm}/}
\def\omit#1{_{\!\rlap{$\scriptscriptstyle \backslash$}
{\scriptscriptstyle #1}}}
\def\vec#1{\mathchoice
        {\mbox{\boldmath $#1$}}
        {\mbox{\boldmath $#1$}}
        {\mbox{\boldmath $\scriptstyle #1$}}
        {\mbox{\boldmath $\scriptscriptstyle #1$}}
}

\begin{abstract}
 We give the alternative formulation of quasiparticle model of quark gluon plasma with medium dependent dispersion relation. The model is thermodynamically consistent provided the medium dependent contribution to the energy density is taken in to account. We establish the connection of our model with other variants of quasiparticle models which are thermodynamically consistent. We test the model by comparing the equation of state with the lattice gauge theory simulations of SU(3) pure gluodynamics .
\end{abstract}


\maketitle

\section{Introduction}
Quasiparticle model of quark gluon plasma is one of the efficient way to characterize the equation of state (EoS); interrelation of state variables of the matter in local thermodynamic equilibrium. The essential feature of this model is that the interacting gas of massless quarks and gluons can be effectively described by ideal gas of massive quarks and gluons\cite{goloviznin}. This behavior is the upshot of refractive nature of the medium so that all the excitations change their dispersion relation. Massless excitations, in particular, becomes massive. This mass which has dynamical origin depends on thermal parameters, temperature (T) and chemical potential ($\mu$) as well as momentum ($\bf{k}$). Such quasiparticle models has been used for studying variety of phenomena such as screening mass of heavy quark potential, transport coefficients like shear and bulk viscosities of pure gluon plasma near QCD phase transition. It has also been used to study the photon and dilepton production rates in heavy-ion collision experiments (HIC).  

The important test of this model is to reproduce the  equation of state as described by lattice quantum chromodynamics (LQCD) simulations. Albeit early phenomenological quasiparticle models\cite{goloviznin,peshierSoff} were able to reproduce LQCD simulation data the issue of thermodynamical consistency of the model kept untouched. This issue was first raised by Gorenstein and Yang\cite{gorenyang} where they pointed out that when the quasiparticle dispersion relation is medium dependent the fundamental thermodynamical relations are not satisfied. In order to deal with this issue they introduced the effective Hamiltonian for the system of quasiparticle quarks and gluons where the ideal gas Hamiltonian is supplemented with the medium dependent vacuum energy term so as to ensure the thermodynamic consistency. This model was further shown to reproduce LQCD simulation data for pure SU(3) gluodynamics\cite{karshGluo}. The issue of thermodynamic consistency was further studied by Bannur\cite{bannur} where the author gave alternative formulation of the quasiparticle model. This work is interesting in the sense that the author was able to recover thermodynamical consistency within standard statistical mechanics without explicit introduction of medium dependent vacuum term. The model is based on so called q-potential which is nothing but the logarithm of the partition function. It can be shown that if one starts with q-potential and follow the standard statistical mechanics method to obtain the pressure, the upshot of medium dependent dispersion relation is to contribute extra term to the pressure itself. This extra term in the pressure together with standard ideal gas expression for energy density ensures the thermodynamical consistency. The model was further shown to explain LQCD simulation data as well.

In this paper we give an alternative formulation of the quasiparticle model which is thermodynamically consistent. We shall see that if the quasiparticle dispersion relation is medium dependent then the energy density picks an extra medium dependent term. This energy density together with ideal gas pressure constitute thermodynamically consistent quasiparticle model. Our model is similar to quasiparticle model of Ref. \cite{bannur} in the sense that it is thermodynamically consistent without explicit contribution of medium dependent vacuum energy of Ref. \cite{gorenyang}, but its different from Ref. \cite{bannur} in the sense that it is the energy density rather than the pressure which picks extra medium dependent term to ensure the thermodynamic consistency. Further we test our model by comparing the EoS with the LQCD simulation results of pure SU(3) gluodynamics. We must emphasize that our model is just an alternative formulation of the quasiparticle model which ensures the thermodynamic consistency when the quasiparticle dispersion relation is medium dependent. The other variants of the quasiaprticle model are as valid as ours as far as the test of the model based on its agreement with LQCD simulation results are concerned.

 We organize the paper as follows. In Sec. II we review the standard statistical mechanics. In Sec. III we describe our quasiparticle model and also show its thermodynamic consistency. In Sect. IV we establish the connection of our model with two other quasiparticle model which are thermodynamically consistent when the dispersion relation are medium dependent. In Sec. V we give the results for the equation of state for quark gluon plasma and compare it with lattice gauge theory simulations of pure SU(3) gluodynamics. Finally in Sec. VI we summarize and conclude. 
\section{Review of statistical mechanics}
Thermodynamical system in which energy ($E_{r}$) as well as number of particles $(N_{s})$ of the subsystems fluctuate about mean values but temperature (T) and chemical potential ($\mu$) are fixed. Such system can be described by the grand canonical ensemble (GCE) where the fundamental quantity from which whole thermodynamics can be derived is the grand canonical partition function $\mathcal{Z}$ given by\cite{landauSM}

\be
\mathcal{Z}=\sum_{s,r}e^{-\beta E_{r}-\alpha N_{s}}=\text{Tr}(e^{-(H-\mu N)/T})
\ee

where $H$ is the Hamiltonian of the system and $\alpha=-\mu/T$, $\beta=1/T$. The thermodynamical potential in this case is the grand potential which is related to the partition function as

\be
\Omega(V,T,\mu)=-T\:\text{ln}\:\mathcal{Z}
\label{GP}
\ee
 The grand potential can also be written as\cite{landauSM}
 
 \be
 \Omega(V,T,\mu)=U-TS-\mu N=-PV
 \label{GPtherm}
 \ee
where P, U, S and N are pressure, internal energy, entropy and number of particles respectively. Then using Eqs. (\ref{GPtherm}) one can obtain macroscopic thermodynamical quantities in terms of grand potential and hence the partition function by taking appropriate derivatives.

\be
P=\frac{T}{V}\text{ln}\:\mathcal{Z}
\label{predef}
\ee
\be
U=-\bigg(\frac{\partial}{\partial \beta}\text{ln}\:\mathcal{Z}\bigg)_{V,\mu}+\frac{\mu}{\beta}\bigg(\frac{\partial}{\partial \mu}\text{ln\:}\mathcal{Z}\bigg)_{V,\beta}
\label{enerdef}
\ee

\be
N=T\bigg(\frac{\partial}{\partial \mu}\text{ln}\:\mathcal{Z}\bigg)_{V,T}
\label{numdef}
\ee

\be
S=\text{ln}\:\mathcal{Z}+T\bigg(\frac{\partial}{\partial T}\text{ln}\:\mathcal{Z}\bigg)_{V,\mu}
\label{entrdef}
\ee

If $H$ is Hamiltonian of the system then one can express above thermodynamical quantities in terms of trace over the eigenstates of $H$ as

\be
P(T,\mu)=\frac{T}{V}\:\text{ln}\: \text{Tr}\:(e^{-(H-\mu N)/T})
\label{pre_stat}
\ee
\be
U/V=\epsilon(T,\mu)=\frac{1}{V}\frac{1}{\mathcal{Z}}\:\text{Tr}(H\: e^{-(H-\mu N)/T})
\label{en_stat}
\ee

\be
N/V=n(T,\mu)=\frac{1}{V}\frac{1}{\mathcal{Z}}\:\text{Tr}(N\: e^{-(H-\mu N)/T})
\ee

 For the case of ideal gas with medium independent dispersion relation it can be shown that pressure and energy density are given by (in the limit $V\longrightarrow \infty$)
\be
P_{id}=\mp d\:T\int_{0}^{\infty}\frac{d^{3}\bf{k}}{(2\pi)^{3}}\text{ln}[1\mp e^{-(\omega-\mu)/T}]
\label{pre_ideal}
\ee
\be
\epsilon_{id}=d\int_{0}^{\infty}\frac{d^{3}\bf{k}}{(2\pi)^{3}}\frac{\omega}{e^{(\omega-\mu)/T}\mp 1}
\label{en_ideal}
\ee

where $\omega=\sqrt{k^{2}+m^{2}}$ is the free particle medium independent dispersion relation and $d$ is the degeneracy factor. Note that two definitions of energy density, $viz.$, Eqs.(\ref{enerdef}) and (\ref{en_stat}) are consistent with each other. It is straightforward to show that these ideal gas pressure and energy density expressions satisfy fundamental thermodynamical relation,
\be
\epsilon(T,\mu)=T\frac{dP}{dT}-P(T,\mu)
\label{therm_rel}
\ee
Thus ideal gas of particles with medium independent dispersion relation is thermodynamically consistent. 

\section{ Statistical mechanics with medium dependent dispersion relation }
Now consider medium dependent dispersion relation which arises solely due to medium dependence of quasiparticle mass.
\be
\omega^{*}(T,\mu)=\sqrt{k^{2}+m^{2}(T,\mu)}
\label{therm_dr}
\ee
Let us derive the expressions for pressure and energy density for the case when $\omega=\omega^*(k,m(T))$. We set $\mu=0$ for simplicity but the generalization for $\mu\neq 0$ is straightforward. In this case the Hamiltonian of the system is medium depenedent.  We observe that the expression for pressure won't change since it does not involve manipulation of $H$ which is T (and $\mu$) dependent. Thus the pressure is given by

\be
P(T)=\frac{T}{V}\:\text{ln}\sum_{r}e^{-E_{r}(T)/T}
\label{pre1}
\ee

 With energy density definition  (\ref{en_stat}) one can obtain 

\be
\epsilon(T)=\frac{1}{V}\langle E_{r}\rangle
\label{ener2}
\ee
where symbol $\langle\rangle$ stands for the thermal average of the quantity inside bracket.  Note that energy density  (\ref{ener2}) together with  pressure given by Eq. (\ref{pre1}) do not satisfy Eq. (\ref{therm_rel}). Hence such quasiparticle model is thermodynamically inconsistent.

But with energy density definition (\ref{enerdef}) one can obtain\cite{banijmp}
\be
\epsilon(T)=\frac{1}{V}\langle E_{r}\rangle+\frac{\beta}{V} \bigg\langle\frac{\partial E_{r}}{\partial \beta}\bigg\rangle
\label{ener1}
\ee

Second term in above equation arises due to medium dependence of dispersion relation. It is straightforward to show that Eqs. (\ref{pre1}) and (\ref{ener1}) satisfy fundamental thermodynamic relation given by Eq. (\ref{therm_rel}) as follows.
\bea
T\frac{\partial P}{\partial T}-P&=&\frac{T}{V}\:\text{ln}\sum_{r}e^{-E_{r}(T)/T}+\frac{1}{V}\frac{\sum_{r}E_{r}e^{-E_{r}(T)/T}}{\sum_{r}e^{-E_{r}(T)/T}}-\frac{T}{V}\frac{\sum_{r}\frac{\partial E_{r}}{\partial T}e^{-E_{r}(T)/T}}{\sum_{r}e^{-E_{r}(T)/T}}-\frac{T}{V}\:\text{ln}\sum_{r}e^{-E_{r}(T)/T}\nonumber\\&=&\frac{1}{V}\langle E_{r}\rangle+\frac{\beta}{V} \bigg\langle\frac{\partial E_{r}}{\partial \beta}\bigg\rangle\nonumber\\&=&\epsilon(T)
\eea

Thus the ideal gas model with medium dependent dispersion relation is also thermodynamically consistent provided the energy density picks an additional term due to medium dependence of dispersion relation. 

In order to figure out the cause of apparent contradiction in the two definitions of energy density, $viz.$, Eq. (\ref{enerdef}) and Eq. (\ref{en_stat}), let us write down the Hamiltonian of the system. When $H$ is medium dependent we can write
\be
H(T,\mu)=H_{id}+\phi(T)
\ee

Where $H_{id}$ is just contribution of ideal gas of quasiparticle excitations and $\phi(T,\mu)$ is the contribution due to thermal medium. This term is similar to mean field potential energy term in the theory of nuclear matter\cite{bugaev} and it can be given the interpretation in terms of vacuum energy\cite{gorenyang}. We assume that $\phi(T,\mu)$ is just a medium effect and it vanishes when dispersion relation is medium independent. With this Hamiltonian the energy density obtained from Eq. (\ref{en_stat}) can be written as
\be
\epsilon_{eff}(T)=\epsilon_{id}(T)+\lim_{V\rightarrow \infty}\frac{\phi(T)}{V}
\label{en_mf}
\ee

The pressure of the system is
\be
P_{eff}(T)=P_{id}(T)-\lim_{V\rightarrow \infty}\frac{\phi(T)}{V}
\label{pre_mf}
\ee

It is straightforward to show that Eqs. (\ref{en_mf}) and (\ref{pre_mf}) obey thermodynamic relation (\ref{therm_rel}) provided $(\frac{\partial P}{\partial m})_{T}=0$ which leads to 

\be
\phi(T)=\phi_{0}-g\int_{T_{0}}^{T}dT^{'}\:m\:\frac{dm}{dT^{'}}\int\frac{d^3\bf{k}}{(2\pi)^3}\frac{1}{\omega^*(k,T^{'})}\frac{1}{e^{\omega^*(k,T^{'})/T^{'}}\pm1}
\label{cons_cond3}
\ee

Energy density obtained from Eq. (\ref{enerdef}) can be written as
\be
\epsilon_{eff}^{'}(T)=\epsilon_{id}+\lim_{V\rightarrow \infty}\frac{\phi}{V}+\lim_{V\rightarrow \infty}\frac{\beta}{V} \bigg\langle\frac{\partial E^{id}_{r}}{\partial \beta}\bigg\rangle+\lim_{V\rightarrow \infty}\frac{\beta}{V} \bigg\langle\frac{\partial \phi}{\partial \beta}\bigg\rangle
\label{eff_en}
\ee

Note that $P_{eff}$ and $\epsilon_{eff}^{'}$ are thermodynamically consistent without extra condition on $\phi$. To make two expressions for energy density, Eqs. (\ref{en_mf}) and (\ref{eff_en}), consistent with each other we set last two terms in Eq. (\ref{eff_en}) equal to zero. Thus we get

\be
\frac{1}{V}\frac{\partial \phi}{\partial \beta}=- \frac{1}{V}\langle\frac{\partial E^{id}_{r}}{\partial \beta}\rangle
\ee
In the limit $V\rightarrow \infty$ above equation can be written after integration as
\be
\phi(T)=\phi_{0}-g\int_{T_{0}}^{T}dT^{'}\:m\:\frac{dm}{dT^{'}}\int\frac{d^3\bf{k}}{(2\pi)^3}\frac{1}{\omega^*(k,T^{'})}\frac{1}{e^{\omega^*(k,T^{'})/T^{'}}\pm1}
\ee
which is same as consistency condition (\ref{cons_cond3}).

Thus we see that energy density definitions  (\ref{en_stat}) and (\ref{enerdef}) are consistent with each other if the contribution $\phi$ is taken into account. Further  $\phi$ should simultaneously satisfy the condition (\ref{cons_cond3}). We conclude that (\ref{enerdef}) is thermodynamically consistent with as well as without consideration of  $\phi$ while Eq. (\ref{en_stat}) is thermodynamically consistent only if we consider contribution of $\phi$ and then putting explicit condition on $\phi$ given by Eq. (\ref{cons_cond3}). 

The requirement of medium dependent contribution apart from single particle energy states and then putting extra conditions on it before using energy density definition (\ref{en_stat}) can be explained if one note that the energy density of the system is related to the pressure and its derivatives through fundamental thermodynamical relation (\ref{therm_rel}). Thus once the pressure is known the energy density is fixed by this basic relation. Now this energy density can be alternatively obtained from Eq. (\ref{en_stat}) and it turns out to be consistent with the energy density obtained from fundamental thermodynamical relation only if the Hamiltonian is medium independent. But we have shown that if one obtain the energy density starting from definition (\ref{enerdef}) it turns out to be consistent with the energy density obtained from fundamental thermodynamical relation even when the Hamiltonian is medium dependent. Thus if one insist that the energy density obtained from the fundamental thermodynamic relation is the true energy density of the system one needs to use Eq. (\ref{enerdef}) as an alternative definition of the energy density rather than Eq. (\ref{en_stat}) to ensure the thermodynamical consistency.

\section{Other quasiparticle models with medium dependent dispersion relation}
Let us discuss other quasiparticle models with medium dependent dispersion relation and the origin of apparent inconsistencies in them. In Ref. \cite{Golo} non-ideal effects were included in the ideal gas model through medium dependent dispersion relation (Eq. (\ref{therm_dr})). Inclusion of non-ideal effects are necessary in order to reproduce LQCD simulation results of interaction measure, $(\epsilon-3P)/T^{4}$. The model is then defined by Eqs. (\ref{en_ideal}) and (\ref{pre_ideal}) for energy density and pressure respectively. 

\subsection{Bag like effective Hamiltonian model}
In Ref. \cite{gorenyang} it was pointed out that such ideal gas model with medium dependent dispersion relation is thermodynamically inconsistent since it violates the fundamental thermodynamical relation given by Eq. (\ref{therm_rel}). The authors argued that such models do not corresponds to quasiparticle models as one can  construct two different ideal gas formulations with different $m(T)$ functions for the same LQCD simulation data; in one formulation one can start with ideal gas pressure given by Eq. (\ref{pre_ideal}) and then calculate energy density from the fundamental thermodynamic relation given by Eq. (\ref{therm_rel}), while in the alternative formulation one starts with ideal gas energy density given by Eq. (\ref{en_ideal}) and then obtain pressure by integrating Eq. (\ref{therm_rel}). In order to remove this inconsistency authors further pointed out that the origin of inconsistency lies in the fact that the Hamiltonian for the system with medium dependent dispersion relation is also medium dependent. The derivation of ideal gas expressions are based on ideal gas Hamiltonian 
\be
H_{id}=\sum_{i=1}^{g}\sum_{\bf{k}}\omega(k)a^{\dag}_{{\bf{k}},i}a_{{\bf{k}},i}
\ee
where the index $i$ corresponds to the particle internal quantum number. Using this Hamiltonian in the statistical mechanics definitions of pressure and energy density given by Eqs. (\ref{pre_stat}) and (\ref{en_stat}) one can obtain ideal gas expressions for pressure and energy density (Eqs. (\ref{pre_ideal}) and (\ref{en_ideal})). But since medium dependence of the Hamiltonian has been ignored energy density thus obtained is inconsistent with the fundamental thermodynamical relation. Remedy to this problem is to include medium dependent zero point energy ($E_{0}^{*}$) explicitly in the Hamiltonian which is otherwise subtracted from the system energy spectrum when the dispersion relation is medium independent. Thus Hamiltonian is rewritten as\cite{gorenyang}
\be
H_{eff}=\sum_{i=1}^{d}\sum_{\bf{k}}\omega(k)a^{\dag}_{{\bf{k}},i}a_{{\bf{k}},i}+E_{0}^{*}
\label{heff}
\ee
With this effective Hamiltonian the fundamental thermodynamical quantities are satisfied with additional condition
\be
\bigg(\frac{\partial P}{\partial m}\bigg)_{T,\mu}=0
\label{cons_cond}
\ee
The pressure and energy density are then
\be
P_{eff}(T,\mu)=\mp d\:T\int_{0}^{\infty}\frac{d^{3}\bf{k}}{(2\pi)^{3}}\text{ln}[1\mp e^{-(\omega^*-\mu)/T}]-B^{*}
\label{pre_eff}
\ee
\be
\epsilon_{eff}(T,\mu)=d\int_{0}^{\infty}\frac{d^{3}\bf{k}}{(2\pi)^{3}}\frac{\omega}{e^{(\omega^*-\mu)/T}\mp 1}+B^{*}
\label{en_eff}
\ee
 where second term in above equations is just medium contribution defined as
 \be
 B^*=B^*(m)=\lim_{V\rightarrow\infty}\frac{E_{0}^{*}}{V}
 \ee
 With above expressions for pressure and energy density the thermodynamical consistency requires that condition (\ref{cons_cond}) should be satisfied which leads to 
 \be
 \frac{dB^*}{dm}=-d\:m(T)\int \frac{d^3\bf{k}}{(2\pi)^3}\frac{1}{\omega^{*}(k,T)}\frac{1}{e^{\omega^*(k,T)/T}\pm1}
 \label{cons_cond1}
 \ee
 
 Thus the effective Hamiltonian formulation of quasiparticle model with medium dependent dispersion relation is thermodynamical consistent. Having said this it is important to note that if statistical mechanics definition given by Eq. (\ref{en_stat}) is used it is necessary to take into account medium dependent vacuum energy contribution for the thermodynamic consistency. But as we have shown in the previous section if the definition (\ref{enerdef}) is used the quasiparticle model is thermodynamically consistent with and without vacuum energy contribution. Thus our model has an added advantage that one need not to worry about vacuum energy contribution.

\subsection{q-potential model}
 In Ref. \cite{bannur} author attempted to reformulate the statistical mechanics starting from  q-potential of the system in which the quasiparticle dispersion relation is medium dependent. This model is quite similar to ours except for the fact that it is pressure rather than the energy density which picks an extra term to guarantee the thermodynamic consistency.   The q-potential is defined as
 \be
 q\equiv\text{ln}\mathcal{Z}=\text{ln}\bigg(\sum_{r,s}e^{-\beta E_{r}-\alpha N_{s}}\bigg)
 \ee
 The q-potential is just logarithm of GCE partition function.  Pressure can be obtained by variation in q-potential. 
 \be
 \delta q=\frac{1}{\mathcal{Z}}\bigg[e^{-\beta(E_{r}-\mu N_{s})}(-E_{r}\delta\beta-\beta\delta E_{r}+N\delta(\beta\mu))\bigg]
 \ee
 Above equation can be rewritten in the form
 \be
 \delta(q+\alpha \langle N\rangle+\beta\langle E\rangle)=\beta\langle \delta Q\rangle-\beta\bigg\langle\frac{\partial E_{r}}{\partial T}\bigg\rangle dT
 \ee
 where we have used the first law of thermodynamics, $\langle \delta Q\rangle=\langle dE\rangle+\langle \delta W\rangle-\mu\langle dN\rangle$. Integrating we get
 \be
 q+\alpha \langle N\rangle+\beta\langle E\rangle=\int\beta\langle \delta Q\rangle-\int\beta\bigg\langle\frac{\partial E_{r}}{\partial \beta}\bigg\rangle d\beta
 \ee
 First term on the right hand side of above equation can be identified with the entropy, $S$. Thus after some manipulation we get
 
 \be 
 \frac{TS+\mu \langle N\rangle-\langle E \rangle}{T}=q+\int\beta\langle\frac{\partial E_{r}}{\partial \beta}\rangle d\beta
 \label{ban1}
 \ee
 
 Using thermodynamic relations $\Omega=\mu\langle N\rangle$ and $\Omega=U-TS+PV$ we get
 \be
 \frac{PV}{T}=q+\int\beta\langle\frac{\partial E_{r}}{\partial \beta}\rangle d\beta
 \label{ban2}
 \ee
  
 \be
 \frac{PV}{T}=\text{ln}\sum_{r,s}e^{-\beta(E_{r}-\mu N_{s})}+\int\: d\beta\:\beta\frac{\partial m}{\partial \beta}\langle\frac{\partial E_{r}}{\partial
 m}\rangle
 \label{pre_ban}
 \ee
 Second term in the above equation arises due to $T$ dependence of quasiparticle mass.
 The energy density in this model is just ideal gas energy density
 \be
 \epsilon=\frac{1}{\mathcal{Z}}\sum_{r,s}E_{r}e^{-\beta E_{r}-\alpha N_{s}}
 \label{en_ban}
 \ee
 It is straightforward to show that Eqs. (\ref{pre_ban}) and (\ref{en_ban}) satisfy fundamental thermodynamical relation (\ref{therm_rel}) and hence model is thermodynamically consistent. This quasiparticle model is similar to ours in the sense that it is thermodynamically consistent without explicit mean field contribution. The only difference is that it is the pressure rather than the energy density which picks extra term to ensure thermodynamic consistency.

\section{Quark gluon plasma EoS}
The essential feature of the quasiparticle model is that the system of interacting particle can be described as a non interacting system of quasiparticles. The effect of the interaction is to modify the dispersion relation of the quasiparticle excitation through their medium dependent masses. In the thermal medium, for momenta $k\sim T$, transverse gluons and quarks propagate with on-shell dispersion relation

\be
\omega^{*}\approx \sqrt{k^2+m^{*2}(T,\mu)}
\ee
where
\be
m^{*2}(T,\mu)=m_{0}^{2}+\Sigma^{*}(T,\mu)
\ee

Neglecting sub leading effects $\Sigma^{*}(T,\mu)$ correspond to leading order on-shell self energies of quarks and gluons in the thermal medium. In the hard thermal loop approximation (HTL)\cite{kapusta,ballac} quark self energy can be obtained as
\be
\Sigma_{q}^{*}(T,\mu)=2\xi_{q}(m_{q0}+\xi_{q}),\: \xi^{2}_{q}(T,\mu)=\frac{N_{c}^{2}-1}{16N_{c}}\bigg\{T^{2}+\frac{\mu_{q}}{\pi^{2}}\bigg\}G^{2}(T,\mu)
\ee
where $m_{0q}$ is the current quark masses and $N_{c}$ is the number of color charges. The gluon self energy in the HTL approximation can be written as
\be
\Sigma_{g}^{*}(T,\mu)=\frac{1}{6}\bigg\{\bigg(N_{c}+\frac{1}{2}N_{f}\bigg)T^{2}+\frac{3}{2\pi^2}\sum_{q}\mu_{q}^{2}\bigg\}G^{2}(T,\mu)
\ee
The entity that is needed to obtain the QGP EoS is effective coupling $G^{2}(T,\mu)$. The {\it{ansatz}} frequently used in the literature is\cite{peshier}
\be
G^{2}(T)=\frac{48\pi^2}{(11N_{c}-2N_{f})\text{ln}\bigg(\lambda\frac{T+T_s}{T_c}\bigg)^{2}}
\ee
The parameter $T_s/T_c$ characterize the exact shape of the quasiparticle spectral function while $T_c/\lambda$ is related to QCD scale $\Lambda_{QCD}$. Above {\it{ansatz}} ensures the deviation from  the exact quasiparticle spectral function  in the thermal medium. Further it approaches the running QCD coupling at high temperatures.

\begin{figure}[h]
\vspace{-0.4cm}
\begin{center}
\begin{tabular}{c c}
 \includegraphics[width=8cm,height=8cm]{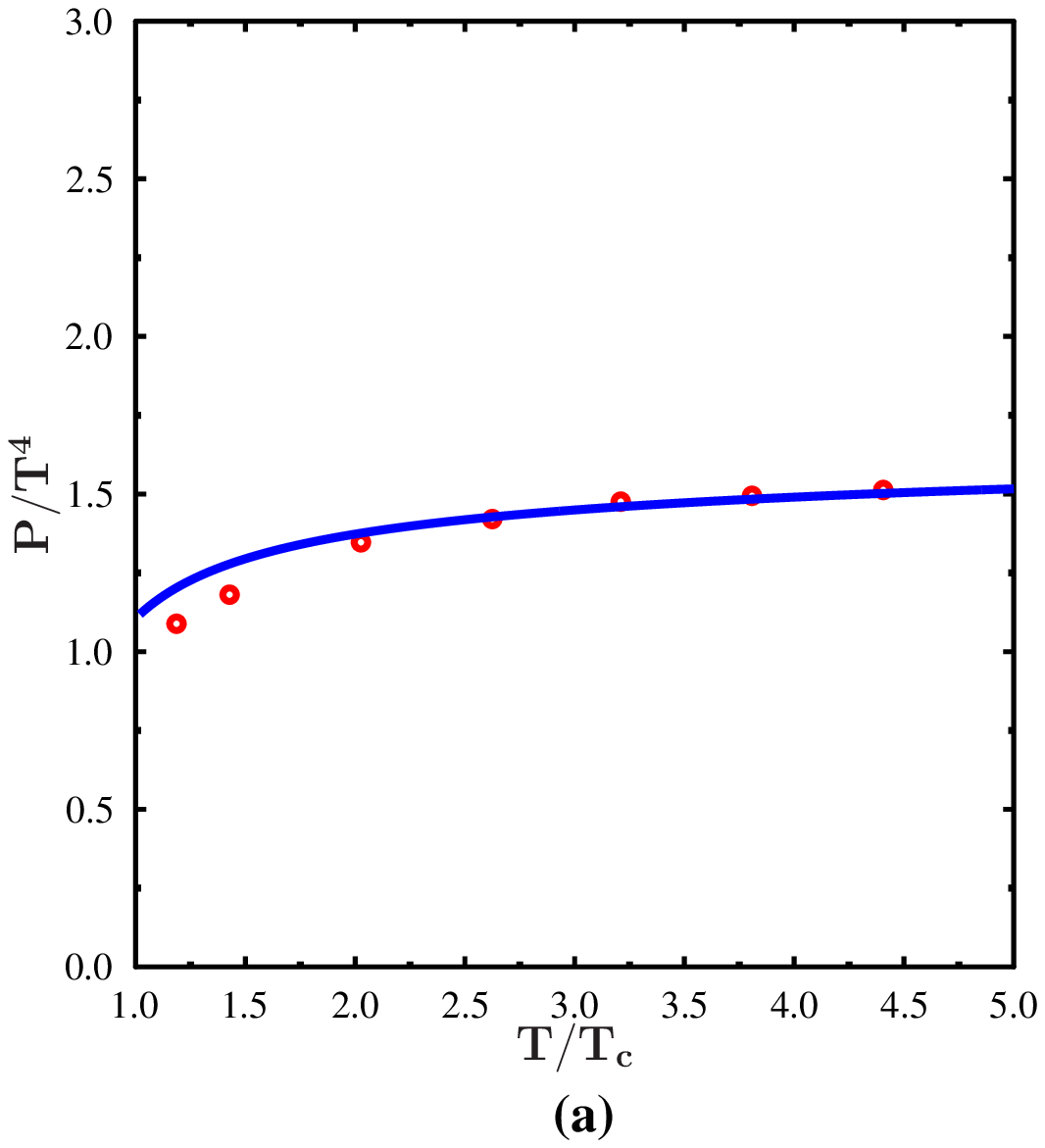}&
  \includegraphics[width=8cm,height=8cm]{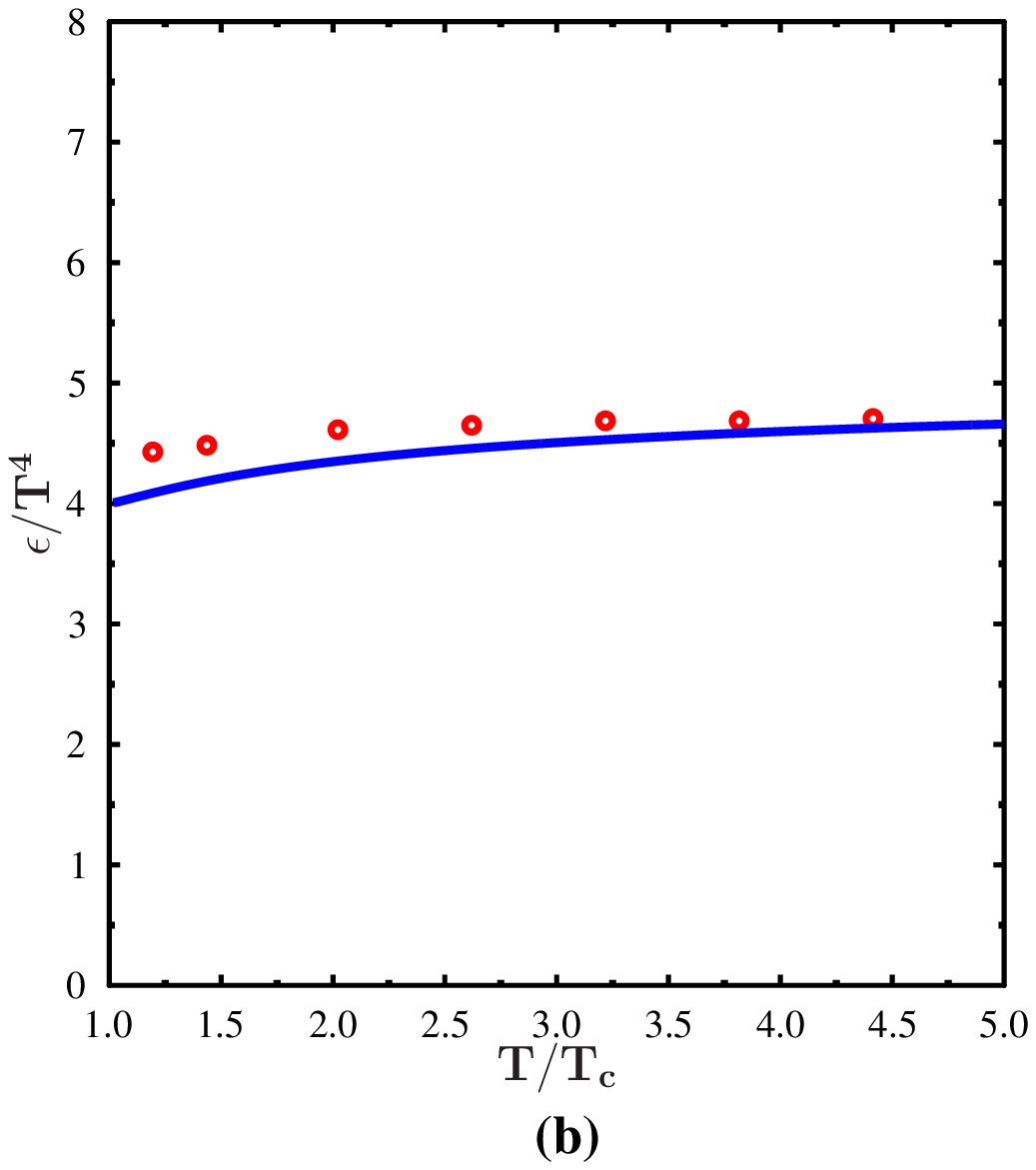}\\
  \end{tabular}
  \caption{(Color online)Left panel shows normalized pressure of pure gluon plasma as a function of temperature (solid blue curve). Right panel shows normalized energy density of the pure gluon plasma (solid blue curve). Red dots corresponds to LQCD simulations of pure SU(3) gluodynamics\cite{karshGluo}.} 
\label{EoS_gluon}
  \end{center}
 \end{figure}

\begin{figure}[h]
\vspace{-0.4cm}
\begin{center}
 \includegraphics[width=8cm,height=8cm]{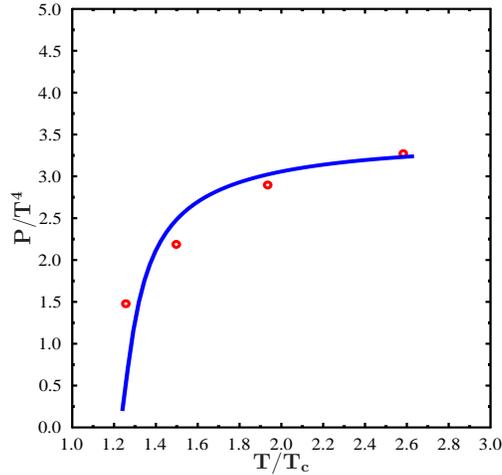}
  \caption{(Color online)Normalized pressure estimation in our model compared with two flavor LQCD simulations\cite{karshtwof}.} 
\label{pre_2f}
  \end{center}
 \end{figure}

Fig. (\ref{EoS_gluon}) shows normalized pressure and energy density of pure gluonic matter estimated in our quasiparticle model. Solid blue curve represents our results while red dots represents lattice QCD simulation of pure SU(3) gluodynamics\cite{karshGluo}. Apart from two fitting parameters  $T_s/T_c$ and $T_c/\lambda$, the gluon degrees of freedom  is also adjusted to fit the lattice data (see table \ref{parameters}). Fig. (\ref{pre_2f}) shows pressure of two flavor quark gluon matter estimated in our model. We observe that with the proper choice of fitting parameters our model is in reasonable agreement with LQCD simulations.

\begin{table}
\begin{center}
    \begin{tabular}{ | l | l | l |p{0.5cm} |}
    \hline
     &  $ \ \ \ \lambda$ & \ \ \  $T_s/T_c$ & $d_{g} $ \\ \hline
    \ \ \   SU(3) & \ \ 5.7 & \ \ -0.4 & \ 18 \\ \hline 
    \ \ \   $N_{f}=2$  & \ \ 2.5 & \ \ -0.8 &\ 16 \\ \hline   
    \end{tabular}
      \caption{The quasiparticle model parameters adjusted to fit the lattice QCD data.}  
      \label{parameters}
\end{center}
\end{table}

 It is important to note that other quasiparticle models we discussed in this paper also reproduce corresponding LQCD data with the proper choice of fitting parameters. Thus it is not possible to decide which model is better over the other merely on the basis of equation of state of that model. Thus the importance of our work lies in the alternative formulation of the quasiparticle model of quark gluon plasma with medium dependent dispersion relation and which is simultaneously thermodynamically consistent. Further unlike quasiparticle model of Ref. \cite{gorenyang} we do not need to supplement pressure and energy density with vacuum energy term which further need to satisfy some constraint in order to ensure the thermodynamic consistency.

 \newpage
\section{Summary and conclusion}
In this paper we have given the alternative formulation of the quasiparticle model of quark gluon plasma with medium dependent dispersion relation. The model is thermodynamically consistent if the extra contribution to the energy density due to medium dependence of quasiparticle mass is taken in to account. We discussed the connection of our model with other thermodynamically consistent quasiparticle models. The bag like quasiparticle model is thermodynamically consistent when the ideal gas expressions for pressure and energy density are supplemented with medium dependent vacuum energy. We found the connection of this model with our model by figuring out the fact that two models use different definitions of the energy density. We showed that our model is thermodynamically consistent with and without inclusion of medium dependent vacuum energy. On the other hand, the q-potential model, being thermodynamically consistent, is similar to our model. Both our model and q-potential model are thermodynamically consistent with and without inclusion of medium dependent vacuum energy. Finally we tested our model by confronting the equation of state with the LQCD simulation results of SU(3) pure gluodynamics and we found reasonable agreement with it.

 
\def\landauSM{L.D. Landau and E.M. Lifshitz, {\it{Statistical Physics}} (Pergamon, Oxford, 1975).}
\def\Golo{V. Goloviznin and H. Satz, Z. Phys. C {\bf 57}, 671 (1993).}
\def\gorenyang{M. I. Gorenstein and S. N. Yang, Phys. Rev. D {\bf 52}, 5206 (1995).}
\def\bannur{V. M. Bannur, Phys. Lett. B {\bf 647}, 271 (2007).}
\def\banijmp{V. M. Bannur, Int. J. Mod. Phys. A {\bf 28}, 130006 (2013).}
\def\bugaev{K. A. Bugaev and M. I. Gorenstein, Z. Phys. C {\bf 43}, 261 (1989).}
\def\karshGluo{G. Boyd, J. Engels, F. Karsch, E. Laermann, C. Legeland, M.
Lu ̈tgemeier, and B. Petersson, Phys. Rev. Lett. {\bf 75}, 4169 (1995)͒. }
\def\karshtwof{F. Karsch, hep-lat/9909006.}
\def\peshier{A. Peshier, B. K\"{a}mpfer and G. Soff, Phys. Rev. C {\bf 61}, 045203 (2000).}
\def\goloviznin{V. Goloviznin and H. Satz, Z. Phys. C {\bf 57}, 671 (1993).}
\def\peshierSoff{A. Peshier, B. K\"{a}mpfer, O. P. Pavlenko and G. Soff, Phys. Lett. B {\bf 337}, 235 (1994).}
\def\kapusta{J. Kapusta and C. Gale, {\it{Finite-Temperature Field Theory: Principles and Applications }} (Cambridge University Press, Cambridge, MA, 2006).}
\def\ballac{M. Le Bellac, {\it{Thermal Field Theory}} (Cambridge University Press, Cambridge, MA, 2000).}

\end{document}